\documentclass[11pt,preprint]{aastex}

\usepackage{longtable}

\newcommand{\etal}{{\it et~al.}}

\begin{document}

\title{Asteroid Polarimetric-Phase Behavior in the Near-Infrared: S- and C-Complex Objects}

\author{Joseph R. Masiero\altaffilmark{1}, S. Tinyanont\altaffilmark{2}, Maxwell A. Millar-Blanchaer\altaffilmark{3}}

\altaffiltext{1}{Caltech/IPAC, 1200 E California Blvd, MC 100-22, Pasadena, CA 91125, USA, {\it jmasiero@ipac.caltech.edu}}
\altaffiltext{2}{University of California, Santa Cruz, CA, 95064, USA}
\altaffiltext{3}{Dept. of Physics, University of California, Santa Barbara, CA, 93106, USA} 

\begin{abstract}

We present the first results of our survey of asteroid
polarization-phase curves in the near-infrared $J$ and $H$ bands using
the WIRC+Pol instrument on the Palomar 200-inch telescope. We confirm
through observations of standard stars that WIRC+Pol can reach the
$0.1\%$ precision needed for asteroid phase curve characterization,
and show that C-complex asteroids could act as an alternate
calibration source as they show less wavelength variation than stellar
polarized standards.  Initial polarization-phase curve results for
S-complex asteroids show a shift in behavior as a function of
wavelength from visible to near-infrared bands, extending previously
observed trends.  Full near-infrared polarization-phase curve
characterization of individual asteroids will provide a unique
constraint on surface composition of these objects by probing the
wavelength dependence of albedo and index of refraction of the surface
material.

\end{abstract}

\section{Introduction}

Light scattered by asteroids encodes information about the physical
makeup of the surface material.  The way that light is polarized as it
is scattered, and the dependence of that polarization on the phase
angle of the observations (the Sun-Asteroid-Observer angle, $\alpha$),
provides unique constraints on the surface properties.  Incoming light
from the Sun incident on the asteroid's surface will be initially
unpolarized.  Simple scattering models predict that the scattered
light will be polarized perpendicular to the scattering plane, with a
maximum near $\alpha\sim90^\circ$.  However observations of
atmosphereless bodies, initially the Moon and then later asteroids,
showed that at phase angles $\alpha<20^\circ$ the observed
polarization direction is reversed and instead shows polarization
parallel to the scattering plane \citep{lyot29,hopfield66,zellner74}.
Detailed scattering models have shown that this reversal of
polarization at small phase angles is due to the Coherent Backscatter
Mechanism that also drives the photometric opposition effect, where
multiply-scattered electromagnetic waves interfere
constructively and destructively in different proportions as a
function of phase \citep{shkuratov94,muinonen02,muinonen09}.

The relationship between the polarization of the scattered light and
phase angle can be described by key quantities that trace physical
surface properties.  For studies of asteroids the angle of
polarization $\theta_r$ is typically measured with respect to the
scattering plane and the polarization $P_r$ is given in this rotated
frame.  Positive $P_r$ values are assigned to polarization
perpendicular to the plane ($\theta_r=0^\circ$) and negative values
assigned to polarization in the plane ($\theta_r=90^\circ$).  The
curve traced by $P_r$ as a function of phase angle $\alpha$ can be
described by four key metrics: 1) the inversion angle $\alpha_0$,
which is the phase angle at which the polarization transitions from
positive to negative; 2) the slope $h$ of the curve at the inversion
angle; 3) the minimum polarization $P_{min}$ reached in the negative
branch; and 4) the phase angle $\alpha_{min}$ at which $P_{min}$ is
reached.  Geometric albedo of the surface has been shown to be
correlated with $h$ and $P_{min}$ \citep{cellino15}, while the
inversion angle is related to the index of refraction of the surface
material \citep{masiero09,gilhutton17}.

These relationships result in the polarization-phase behavior showing
a dependence on the composition of an asteroid's surface, as traced by
its spectral taxonomic classification.  A major survey of visible
light asteroid polarization was carried out using the Complejo
Astron\'{o}mico El Leoncito (CASLEO) facility in Argentina
\citep{gilhutton11,gilhutton12,canada12}, covering a wide range of
taxonomic classes.  The CASLEO results showed that within a taxonomic
class objects tend to show consistent polarimetric behavior, while
between classes changes in $h$, $P_{min}$, $\alpha_{min}$, and
$\alpha_0$ are observed and tend to follow compositional trends.  This
survey has continued in recent years with the aim of probing smaller
objects to search for size-dependent behaviors, correlations with
asteroid family, and end-member taxonomies.

Asteroid polarization should also depend on the wavelength of light
used for the observations, as both reflectivity and index of
refraction can have wavelength dependencies.  While some asteroids,
such as the B- and C-types, have roughly flat spectra from the visible
to the near-infrared, the S-complex objects show a red spectral slope,
and this reflectivity change would be predicted to correspond to a
change in $h$ slope and $P_{min}$.  \citet{belskaya09} obtained
asteroid polarimetric phase curves spanning the UBVRI bandpasses for a
range of asteroid compositions.  They found a wavelength dependence of
a few tenths of a percent polarization per micron across most phase
angles, with S-types at larger phases showing the largest change.
However, contrary to expectations, the sign of the change did not
reverse at the inversion angle, indicating that instead of a pivot
around inversion angle the curves were offset.  In a different survey,
\citet{pan22} used the TriPol instrument to observe polarimetric phase
curves in the $g^\prime$, $r^\prime$, and $i^\prime$ bands, but found
only minimal changes with wavelength.

Up until now, asteroid polarization surveys have focused on
wavelengths shortward of $\sim1~\mu$m.  The first asteroid surveys
were conducted primarily in the B and G bands as they made use of
photomultiplier tubes that were preferentially sensitive to those
wavelengths \citep[e.g.][]{zellner76}.  Later surveys such those using
the CASLEO instrument \citep[e.g.][]{cellino05} spanned the U, B, V,
R, I bands, though the majority of observations were obtained in V and
R where asteroid reflected light flux peaks.  Here we report our first
results from a survey of asteroid polarization-phase curves in the
near-infrared $J$ ($1.25~\mu$m) and $H$ ($1.64~\mu$m) bands.  This is
the first such survey of asteroid polarization at these
wavelengths. It should be noted that there have been previous
investigations of comet polarizations at these wavelengths
\citep[e.g.][]{oishi78,jones08,kwon19} however they generally have
been at much higher phase angles where polarizations are larger, or
have not had sufficient instrumental accuracy to characterize the
negative polarization branch.  In this paper we describe our observing
methodology and present early results from our observations.  Future
work will focus on constraining the polarization-phase behavior of
specific objects as well as compositional groupings and sub-classes
(e.g. the Barbarian asteroids).

\section{Observations}

To carry out our survey, we make use of the newly commissioned
polarimetric mode for the Wide-field Infrared Camera (WIRC) instrument
on the 200-inch Palomar telescope.  WIRC is a near-infrared optimized
camera mounted at the telescope's prime focus, providing an
8.7$^\prime$ wide field of view with $0.25''$ pixels \citep{wilson03}.
The 200-inch is an equatorially-mounted telescope, meaning that WIRC
will maintain an unchanging orientation with respect to the sky and
that there are no fold mirrors in the light path, making it an ideal
choice for adding polarimetric sensitivity.

A polarimetric enhancement to WIRC, called WIRC+Pol, was deployed in
2017 \citep{tinyanont19a} using a novel combination of a quarter-wave
plate in series with polarization grating (PG) to simultaneously
measure the four different components of linear polarization
($0^\circ$, $45^\circ$, $90^\circ$, $135^\circ$; referred to a +Q, +U,
-Q, and -U in the Stokes vector formalism).  The PG element splits the
light into the four polarization components while simultaneously
dispersing each into low-resolution (R$~\sim100$) spectra, providing
spectropolarimetric sensitivity as well.  Because of this 4-way
dispersion, the field of view is restricted by an upstream mask, and
observations of extended objects are more complicated to
untangle. Additionally, only the $J$ and $H$ bands are usable in this
spectropolarimetric mode as much of the field for $K$ band
observations would be dispersed beyond the detector.

In 2019, WIRC+Pol was upgraded with a half-wave plate (HWP) upstream
of the PG \citep{tinyanont19b}.  By rotating the half-wave plate
through four different orientations ($\theta_{hwp}=0^\circ$,
$22.5^\circ$, $45^\circ$, $67.5^\circ$) the polarization of the
incoming light becomes rotated by $2\theta_{hwp}$, effectively
swapping the Stokes parameter that contributes to each of the four
beams leaving the PG.  This beam-swapping allows for slight
differences in the optical paths between the beams to be corrected
for, and pushes the polarimetric precision of WIRC+Pol to $<0.1\%$
when sufficient signal-to-noise is obtained.

Our typical observing sequence for the data presented here consisted
of a series of four cycles through the four HWP angles, for a total of
sixteen exposures before switching to the other filter.  For
observations in 2021, we enhanced this cadence by observing A-B pairs
to better constrain the background for faint targets.  All targets
observed in 2019 were brighter than $J<10.5~$mag and so did not show
any systematic differences due to lack of A-B pairs. In this mode, we
obtained four HWP cycles in $J$ band at position A, four cycles in $H$
band at position A, four cycles in $H$ at position $B$, and finally
four cycle in $J$ at position B.  This 64 image set was our primary
observing unit, and causes the UT midpoint of both bands to be the
same.  With readout, filter change, and dithering overheads this
observing block takes approximately 20 mins of wall-clock time for
bright sources ($J=6-11~$mag, exposure times of 1-5 seconds) and about
50 mins for faint sources ($J=11-13~$mag, exposure time of 30
seconds). For our faintest targets this sequence was repeated as
necessary to obtain sufficient flux for high precision polarimetry.

\section{Data Reduction and Calibration Verification}

All image calibration and polarization measurements were accomplished
using the WIRC+Pol Data Reduction Pipeline available publicly on
GitHub\footnote{\textit{https://github.com/WIRC-Pol/wirc\_drp}}.  This
pipeline makes use the \texttt{Numpy}, \texttt{Scipy},
\texttt{Astropy}, \texttt{PhotUtils}, and \texttt{Matplotlib} Python
packages.  The pipeline extracts polarimetric measurements from each
of the four diffracted beams and combines them over the HWP rotations
to determine the final polarization degree and angle.  The WIRC+Pol
Data Reduction Pipeline is described in \citet{tinyanont19a} and
updated details are available at the documentation
website\footnote{\textit{https://wircpol.readthedocs.io/}}.

In order to calibrate the raw FITS images from WIRC+Pol, we dark subtract
and flatten each individual image before extracting the polarimetric
measurements.  A master dark file was created each night for each
exposure time, comprised of 30 dark frames from the beginning of the
night and 30 from the end.  Dome flats were obtained each night for
both bands, and a master flat field was created with all the optical
elements in the beam.  

For data from our 2019 runs, background levels were interpolated from
nearby regions on the detector.  The observations from 2021 were taken
in an A-B observing sequence and so background levels from the B
frames were subtracted from the same pixels as the location of the
data in the A frames, and vice versa.  As described in
\citet{tinyanont19b}, to correct for beam throughput the differences
for Q and U beam pairs from each waveplate position were calculated
and then averaged over all waveplate orientations.  Because WIRC+Pol
disperses the light in each band, final Q and U values are output as a
function of wavelength.  For this work, we provide the error-weighted
mean of the polarization degree $P$ and polarization angle $\theta$
for wavelengths with sufficient throughput.  This allows us to improve
our signal-to-noise while investigating polarization changes from
visible to $J$ to $H$ bands.  Future work will explore the
polarization as a function of wavelength within each band.

\begin{center}
\scriptsize
  \noindent
\begin{longtable}{cllcllll}
\caption{WIRC+Pol Standard Star Results}\label{tab.stardata}\\
\hline\hline

Target & Date & Midpoint  & Filter & Literature & Literature & Measured & Measured  \\ 
 &  & UT &  & Polarization & Polarization & Polarization & Angle \\ 
 &  &    &  & Value & Angle (deg E of N) &  & (deg) \\ 

\hline\hline
\endfirsthead
\caption[]{(continued)}\\
\hline\hline
Target & Date & Midpoint  & Filter & Literature & Literature & Measured & Measured  \\ 
 &  & UT &  & Polarization & Polarization & Polarization & Angle \\ 
 &  &    &  & Value & Angle (deg E of N) &  &  (deg) \\ 
\hline\hline
\endhead
\hline
\endfoot

\hline
\textbf{Polarized}& & & & & & &\\
\textbf{Standards}& & & & & & &\\
\hline

HD17747   &  &  & V & $0.89\pm0.06\%$ & $139.7\pm1.8$ &  & \\ 
HD17747   & 2021-09-04 & 10:27 & J & $0.41\pm0.06\%$ & $139.7\pm1.8$ & $0.354\pm0.025\%$ & $125.8\pm  3.7$\\ 
HD17747   & 2021-09-04 & 10:27 & H & $0.22\pm0.06\%$ & $139.7\pm1.8$ & $0.234\pm0.009\%$ & $116.1\pm  2.0$\\ 
HD17747   & 2021-11-08 & 07:57 & J & $0.41\pm0.06\%$ & $139.7\pm1.8$ & $0.336\pm0.011\%$ & $124.7\pm  2.2$\\ 
HD17747   & 2021-11-08 & 07:57 & H & $0.22\pm0.06\%$ & $139.7\pm1.8$ & $0.214\pm0.011\%$ & $109.5\pm  1.6$\\ 

HD30870    &  &  & V & $1.34\pm0.2\%$ & $66.0\pm4.3$ &  & \\ 
HD30870    & 2019-08-30 & 11:31 & J & $0.62\pm0.2\%$ & $66.0\pm4.3$ & $0.643\pm0.070\%$ & $ 86.8\pm  2.3$\\ 

HD35395   &  & & V & $1.52\pm0.20\%$ & $147\pm3.8$ &  & \\ 
HD35395   & 2021-02-03 & 03:51 & J & $0.70\pm0.20\%$ & $147\pm3.8$ & $0.572\pm0.010\%$ & $140.6\pm  1.0$\\ 
HD35395   & 2021-02-03 & 03:51 & H & $0.38\pm0.20\%$ & $147\pm3.8$ & $0.293\pm0.002\%$ & $138.2\pm  1.0$\\ 

HD46660   &  & & V & $1.75\pm0.20\%$ & $13\pm3$ &  & \\ 
HD46660   & 2021-02-03 & 05:13 & J & $0.81\pm0.20\%$ & $13\pm3$ & $0.983\pm0.012\%$ & $  9.3\pm  0.2$\\ 
HD46660   & 2021-02-03 & 05:13 & H & $0.44\pm0.20\%$ & $13\pm3$ & $0.713\pm0.011\%$ & $ 12.6\pm  0.2$\\ 

HD50064   &  &  & V & $2.21\pm0.18\%$ & $150\pm2.3$ &  & \\ 
HD50064   & 2019-03-17 & 03:47 & J & $1.02\pm0.18\%$ & $150\pm2.3$ & $1.182\pm0.010\%$ & $144.9\pm  0.5$\\ 
HD50064   & 2019-03-17 & 04:10 & H & $0.55\pm0.18\%$ & $150\pm2.3$ & $0.885\pm0.005\%$ & $145.8\pm  0.9$\\ 
HD50064   & 2019-03-17 & 05:54 & J & $1.02\pm0.18\%$ & $150\pm2.3$ & $1.185\pm0.015\%$ & $144.8\pm  0.7$\\ 
HD50064   & 2019-03-17 & 06:04 & H & $0.55\pm0.18\%$ & $150\pm2.3$ & $0.716\pm0.013\%$ & $149.4\pm  1.0$\\ 

HD58624   &  &  & V & $1.11\pm0.07\%$ & $27\pm1.8$ &  & \\ 
HD58624   & 2019-03-17 & 06:42 & J & $0.51\pm0.07\%$ & $27\pm1.8$ & $0.341\pm0.014\%$ & $ 26.4\pm  1.5$\\ 
HD58624   & 2019-03-17 & 06:23 & H & $0.28\pm0.07\%$ & $27\pm1.8$ & $0.302\pm0.016\%$ & $ 18.9\pm  1.3$\\
HD58624   & 2021-02-03 & 06:48 & J & $0.51\pm0.07\%$ & $27\pm2$ & $0.365\pm0.017\%$ & $ 22.9\pm  1.4$\\ 
HD58624   & 2021-02-03 & 06:48 & H & $0.28\pm0.07\%$ & $27\pm2$ & $0.230\pm0.006\%$ & $ 21.4\pm  1.0$\\
HD58624   & 2021-11-08 & 12:53 & J & $0.56\pm0.07\%$ & $26.9\pm1.8$ & $0.734\pm0.052\%$ & $ 12.6\pm  1.2$\\ 
HD58624   & 2021-11-08 & 12:53 & H & $0.28\pm0.07\%$ & $26.9\pm1.8$ & $0.421\pm0.009\%$ & $ 19.5\pm  1.7$\\ 

HD144639  &  & & V & $1.06\pm0.05\%$ & $87\pm1.3$ &  & \\ 
HD144639  & 2021-06-26 & 06:55 & J & $0.49\pm0.05\%$ & $87\pm1.3$ & $0.593\pm0.023\%$ & $ 77.2\pm  0.7$\\ 
HD144639  & 2021-06-26 & 06:55 & H & $0.26\pm0.05\%$ & $87\pm1.3$ & $0.315\pm0.008\%$ & $ 90.3\pm  0.2$\\ 

HD152466  &  &  & V & $1.01\pm0.00\%$ & $88.9\pm0.4$ &  & \\ 
HD152466  & 2019-03-17 & 11:42 & J & $0.47\pm0.00\%$ & $88.9\pm0.4$ & $0.601\pm0.014\%$ & $ 83.4\pm  0.3$\\ 

HD188618   &  &  & V & $2.40\pm0.20\%$ & $179\pm2.4$ &  & \\ 
HD188618   & 2019-08-30 & 04:34 & J & $1.11\pm0.20\%$ & $179\pm2.4$ & $1.206\pm0.024\%$ & $166.9\pm  0.4$\\ 
HD188618   & 2019-08-30 & 06:30 & J & $1.11\pm0.20\%$ & $179\pm2.4$ & $1.333\pm0.037\%$ & $164.6\pm  0.6$\\ 

HD208205  &  & & V & $0.70\pm0.07\%$ & $145\pm3$ &  & \\ 
HD208205  & 2021-05-30 & 10:39 & J & $0.32\pm0.07\%$ & $145\pm3$ & $0.401\pm0.010\%$ & $135.2\pm  1.7$\\ 
HD208205  & 2021-05-30 & 10:39 & H & $0.18\pm0.07\%$ & $145\pm3$ & $0.313\pm0.004\%$ & $127.4\pm  2.5$\\ 

HD220859  &  &  & V & $0.58\pm0.03\%$ & $125.8\pm1.4$ &  & \\ 
HD220859  & 2021-09-04 & 07:15 & J & $0.27\pm0.03\%$ & $125.8\pm1.4$ & $0.262\pm0.043\%$ & $ 81.2\pm  2.2$\\ 
HD220859  & 2021-09-04 & 07:15 & H & $0.15\pm0.03\%$ & $125.8\pm1.4$ & $0.206\pm0.030\%$ & $ 101.1\pm  1.7$\\ 
HD220859  & 2021-11-08 & 04:04 & J & $0.27\pm0.03\%$ & $125.8\pm1.4$ & $0.236\pm0.035\%$ & $118.4\pm  5.7$\\ 

HD231195  &  & & V & $3.67\pm0.06\%$ & $34.4\pm0.5$ &  & \\ 
HD231195  & 2021-06-26 & 10:26 & J & $1.69\pm0.06\%$ & $34.4\pm0.5$ & $1.250\pm0.121\%$ & $ 44.9\pm  5.2$\\

\hline
\textbf{Unpolarized}& & & & & & &\\
\textbf{Standards}& & & & & & &\\
\hline

HD26515    &  &  & V & $0.02\pm0.05\%$ &  &  & \\ 
HD26515    & 2019-08-30 & 08:54 & J & $0.02\pm0.05\%$ &  & $0.141\pm0.017\%$ & $ 68.5\pm  4.0$\\ 
HD26515    & 2019-08-30 & 11:50 & J & $0.02\pm0.05\%$ &  & $0.198\pm0.030\%$ & $103.3\pm  2.8$\\

HD35076   &  &  & V & $0.07\pm0.03\%$ &  & & \\ 
HD35076   & 2021-09-04 & 12:22 & J & $0.07\pm0.03\%$ &  & $0.125\pm0.022\%$ & $ 41.9\pm  8.6$\\ 
HD35076   & 2021-09-04 & 12:22 & H & $0.07\pm0.03\%$ &  & $0.208\pm0.018\%$ & $ 77.3\pm  2.1$\\ 
HD35076   & 2021-11-08 & 11:10 & J & $0.07\pm0.03\%$ &  & $0.318\pm0.060\%$ & $ 89.9\pm  2.1$\\ 
HD35076   & 2021-11-08 & 11:10 & H & $0.07\pm0.03\%$ &  & $0.257\pm0.006\%$ & $ 71.2\pm  1.4$\\ 

HD40724   &  & & V & $0.05\pm0.03\%$ &  &  & \\ 
HD40724   & 2021-02-03 & 04:13 & J & $0.05\pm0.03\%$ &  & $0.094\pm0.026\%$ & $ 49.2\pm  3.0$\\ 
HD40724   & 2021-02-03 & 04:13 & H & $0.05\pm0.03\%$ &  & $0.217\pm0.007\%$ & $121.6\pm  0.1$\\ 

HD51596   &  &  & V & $0.02\pm0.03\%$ &  &  & \\ 
HD51596   & 2019-03-17 & 04:35 & J & $0.02\pm0.03\%$ &  & $0.074\pm0.013\%$ & $ 46.3\pm  5.0$\\ 
HD51596   & 2019-03-17 & 04:28 & H & $0.02\pm0.03\%$ &  & $0.185\pm0.008\%$ & $114.4\pm  2.9$\\ 
HD51596   & 2019-03-17 & 05:48 & J & $0.02\pm0.03\%$ &  & $0.163\pm0.013\%$ & $ 31.9\pm  3.7$\\ 
HD51596   & 2019-03-17 & 05:36 & H & $0.02\pm0.03\%$ &  & $0.064\pm0.006\%$ & $ 53.0\pm  0.5$\\ 

HD65970   &  &  & V & $0.04\pm0.04\%$ &  &  & \\
HD65970   & 2019-03-17 & 07:05 & J & $0.04\pm0.04\%$ &  & $0.032\pm0.012\%$ & $ 48.1\pm  7.2$\\
HD65970   & 2019-03-17 & 07:28 & H & $0.04\pm0.04\%$ &  & $0.058\pm0.009\%$ & $ 90.4\pm  0.1$\\

HD71371   &  & & V & $0.05\pm0.05\%$ &  &  & \\ 
HD71371   & 2021-02-03 & 06:22 & J & $0.05\pm0.05\%$ &  & $0.080\pm0.007\%$ & $ 44.7\pm  3.8$\\ 
HD71371   & 2021-02-03 & 06:22 & H & $0.05\pm0.05\%$ &  & $0.052\pm0.007\%$ & $ 66.0\pm  2.1$\\ 

HD79096   &  & & V & $0.03\pm0.04\%$ &  &  & \\ 
HD79096   & 2021-02-03 & 07:11 & J & $0.03\pm0.04\%$ &  & $0.031\pm0.006\%$ & $110.1\pm  5.1$\\ 
HD79096   & 2021-02-03 & 07:11 & H & $0.03\pm0.04\%$ &  & $0.080\pm0.007\%$ & $ 43.0\pm  0.8$\\ 
HD79096   & 2021-05-30 & 03:49 & J & $0.03\pm0.04\%$ &  & $0.169\pm0.044\%$ & $ 71.4\pm 10.6$\\ 
HD79096   & 2021-05-30 & 03:49 & H & $0.03\pm0.04\%$ &  & $0.186\pm0.006\%$ & $ 90.3\pm  0.2$\\ 

HD105262  &  & & V & $0.07\pm0.04\%$ &  &  & \\ 
HD105262  & 2021-02-03 & 11:20 & J & $0.07\pm0.04\%$ &  & $0.158\pm0.037\%$ & $ 23.2\pm  2.9$\\ 
HD105262  & 2021-02-03 & 11:20 & H & $0.07\pm0.04\%$ &  & $0.131\pm0.005\%$ & $ 69.9\pm  2.9$\\ 

HD105928  &  & & V & $0.037\pm0.034\%$ &  &  & \\ 
HD105928  & 2021-06-26 & 03:55 & J & $0.037\pm0.034\%$ &  & $0.061\pm0.012\%$ & $ 85.7\pm  4.0$\\ 
HD105928  & 2021-06-26 & 03:55 & H & $0.037\pm0.034\%$ &  & $0.152\pm0.006\%$ & $  0.9\pm  0.2$\\ 

HD136497  &  &  & V & $0.04\pm0.01\%$ &  &  & \\ 
HD136497  & 2019-03-17 & 11:05 & J & $0.04\pm0.01\%$ &  & $0.046\pm0.010\%$ & $ 72.6\pm  5.4$\\ 
HD136497  & 2019-03-17 & 10:43 & H & $0.04\pm0.01\%$ &  & $0.122\pm0.005\%$ & $ 69.9\pm  4.7$\\

HD196348   &  &  & V & $0.06\pm0.04\%$ &  &  & \\ 
HD196348   & 2019-08-30 & 04:49 & J & $0.06\pm0.04\%$ &  & $0.476\pm0.096\%$ & $ 52.9\pm 12.9$\\ 
HD196348   & 2019-08-30 & 06:15 & J & $0.06\pm0.04\%$ &  & $0.458\pm0.121\%$ & $149.9\pm  5.2$\\ 

HD197577  &  & & V & $0.07\pm0.04\%$ &  &  & \\ 
HD197577  & 2021-06-26 & 07:44 & J & $0.07\pm0.04\%$ &  & $0.123\pm0.021\%$ & $ 88.9\pm  2.5$\\ 
HD197577  & 2021-06-26 & 07:44 & H & $0.07\pm0.04\%$ &  & $0.058\pm0.005\%$ & $125.9\pm  1.5$\\ 
HD197577  & 2021-09-04 & 05:46 & J & $0.07\pm0.04\%$ &  & $0.157\pm0.028\%$ & $ 88.9\pm  2.4$\\ 
HD197577  & 2021-09-04 & 05:46 & H & $0.07\pm0.04\%$ &  & $0.076\pm0.007\%$ & $ 90.0\pm  0.1$\\ 
HD197577  & 2021-11-08 & 02:07 & J & $0.07\pm0.04\%$ &  & $0.311\pm0.029\%$ & $151.0\pm  3.6$\\ 
HD197577  & 2021-11-08 & 02:07 & H & $0.07\pm0.04\%$ &  & $0.156\pm0.018\%$ & $ 41.3\pm  2.0$\\ 

HD203843   &  &  & V & $0.06\pm0.04\%$ &  &  & \\ 
HD203843   & 2019-08-30 & 06:59 & J & $0.06\pm0.04\%$ &  & $0.285\pm0.053\%$ & $ 21.9\pm  4.1$\\ 
HD203843   & 2019-08-30 & 08:36 & J & $0.06\pm0.04\%$ &  & $0.252\pm0.069\%$ & $ 68.8\pm  3.2$\\ 
HD203843  & 2021-05-30 & 10:18 & J & $0.06\pm0.04\%$ &  & $0.219\pm0.033\%$ & $118.6\pm  1.8$\\ 
HD203843  & 2021-05-30 & 10:18 & H & $0.06\pm0.04\%$ &  & $0.322\pm0.033\%$ & $113.4\pm  0.4$\\ 

HD221356  &  &  & V & $0.02\pm0.01\%$ &  &  & \\ 
HD221356  & 2021-09-04 & 07:50 & J & $0.02\pm0.01\%$ &  & $0.094\pm0.060\%$ & $ 64.7\pm 12.0$\\ 
HD221356  & 2021-09-04 & 07:50 & H & $0.02\pm0.01\%$ &  & $0.083\pm0.014\%$ & $ 53.6\pm  6.5$\\

\hline
\end{longtable}
\end{center}

\clearpage

Table~\ref{tab.stardata} presents our measurements of the polarized
and unpolarized stars observed as part of our survey.  The
spectropolarimetric measurements from WIRC+Pol were combined,
weighting by their uncertainties, to generate the broad-band
polarization and angles given here.  Literature polarization values
for each object at $V$ band are given, as well as the estimated
polarization for these standards at $J$ and $H$ wavelengths based on
literature measurements (see discussion below).  Angles of
polarization are given in degrees east of north.

In order to validate the calibration of the instrument and data
reduction software, and ensure that there were no unexpected
night-to-night variations, we obtained measurements of polarized and
unpolarized standard stars on each observing night.  Standards were
chosen to be as close as possible on-sky to our asteroid targets, and
were observed such that their hour angle was similar to that of our
asteroid targets.  For our observing dates in 2019 we repeated
standards before and after the targets to constrain any short-term
variations, but our results were repeatable within the expected error
($\Delta P \sim0.1\%$) and so for future observing runs standards were
only observed once per night to increase the time available for
asteroid observations.

Polarized and unpolarized standards were chosen from the
\citet{heiles00} compilation of standard stars for polarimetry.  The
compiled literature measurements were from a variety of source works,
but all were obtained at visible wavelengths. Polarization of stellar
light by interstellar dust, which is the main source of polarization
in bright stars, shows a distinct dependence on wavelength
\citep{serkowski75}.  To estimate the expected polarization level of
our polarized standards at the $J$ and $H$ bands, we use Eq. (4) from
\citet{serkowski75}:
\[P/P_{max} = exp [-1.15 \ln^2(\lambda_{max}/\lambda)]\]
where $P$ is the expected polarization, $P_{max}$ is the maximum
polarization (assumed to be the literature value), $\lambda_{max}$ is
the wavelength of maximum polarization (assumed to be $0.55\mu$m),
$\lambda$ is the wavelength of observation.  For our $J$ and $H$ band
measurements, this results in a depolarization of the standard value
of $0.461$ for $J$ and $0.250$ for $H$.  The \citet{serkowski75}
correction was determined using the $U$, $B$, $V$, and $R$ bandpasses
only, so there is an unknown uncertainty associated with extending
this to $J$ and $H$.  The angle of polarization ($\theta$) in the
near-infrared was assumed to be the same as for the literature
measurements.

\begin{figure}[ht]
\begin{center}
\includegraphics[scale=0.5]{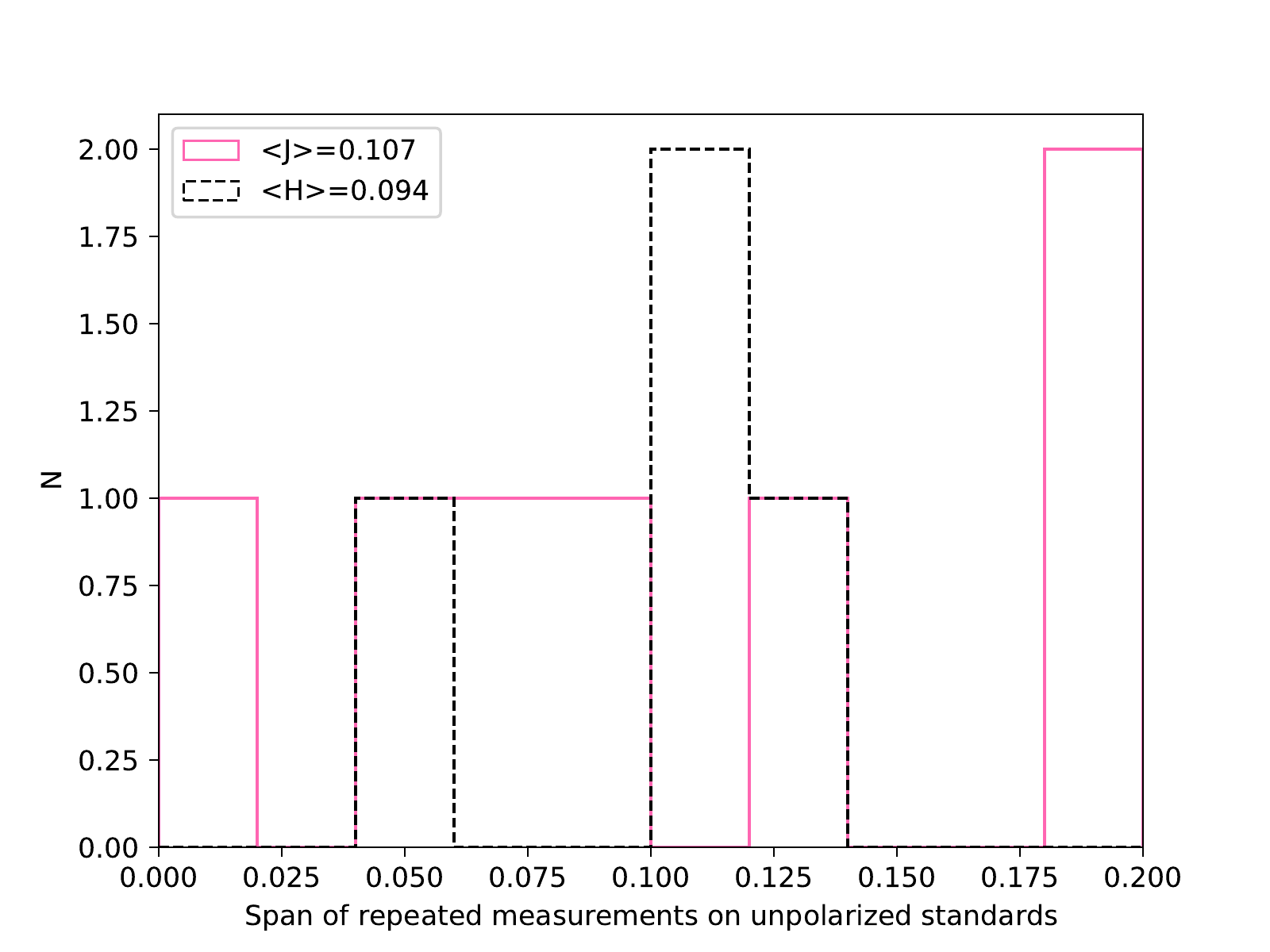}
\protect\caption{Histogram showing the changes in measured
  polarization between observations for nominally unpolarized standard
  stars.  X-axis units are in percent polarization, and the mean for
  each band, given in the legend, is consistent with $0.1\%$ overall
  system precision.}
\label{fig.unpol}
\end{center}
\end{figure}

Our measurements of unpolarized standards generally show absolute
polarization values that scatter around the expected $0.1\%$ level.
However, some standards (e.g. HD35076 and HD203843) repeatedly show
measured polarizations larger than this on different nights.  This
could be due to instrumental polarization.  A more likely explanation
is that these sources are unpolarized at visible wavelengths but have
non-zero polarization in the near-infrared, given their measured
repeatability and the non-detection of polarization for other stars.
We assess the instrumental polarization by looking at the measurement
scatter for unpolarized standards observed multiple times (either
within a run or between runs).  We show in Figure~\ref{fig.unpol} a
histogram of the maximum change in polarization between observations
for multiply-measured unpolarized standards.  The mean of these values
is around $0.1\%$, consistent with the accuracy expected from previous
calibration analyses \citep{tinyanont19b,millarblanchaer21}.

\begin{figure}[ht]
\begin{center}
\includegraphics[scale=0.5]{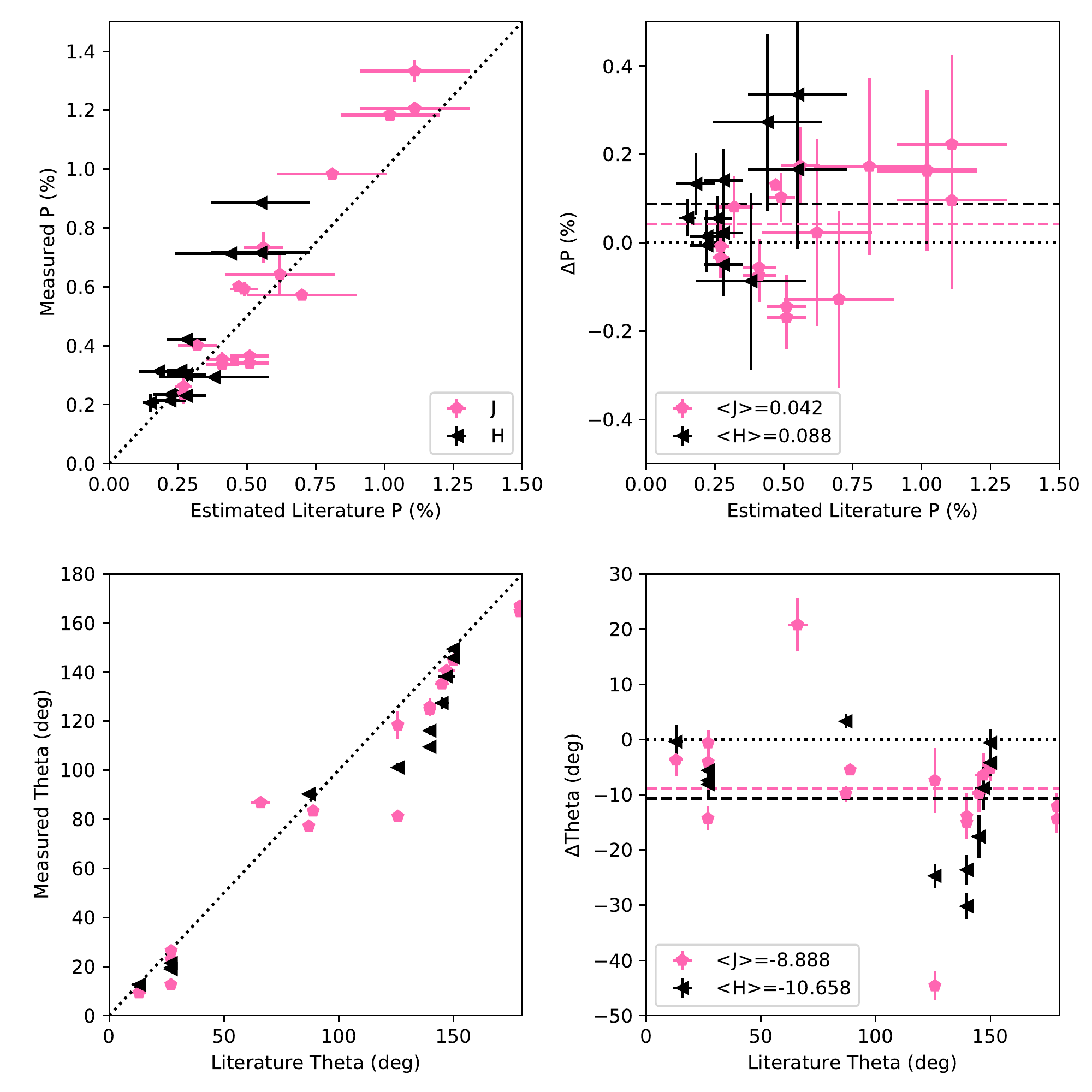}
\protect\caption{Comparison of predicted and measured polarization
  degrees and angles for polarized standard stars. The top-left shows
  the measured polarization value against the literature values
  corrected for wavelength changes. Top-right shows the difference in
  polarization (measured-literature) against the scaled literature
  values.  Bottom-left shows the measured angle of polarization
  compared to the literature angle, and the bottom-right shows the
  difference in angle compared to the literature angle.  Both
  difference plots include dashed lines showing the mean offset
  values, with the value for each band given in the legend.}
\label{fig.pol}
\end{center}
\end{figure}

Observations of polarized standards allow us to check for potential
instrumental depolarization and validate the measured polarization
angle.  We show in Figure~\ref{fig.pol} the results of our comparison
of our measured values to literature values. It is important to note
that because these literature values are scaled via a depolarization
factor to correct for wavelength-dependent polarization, the true
uncertainty on these measurements is certainly larger than the quoted
values that are adopted from the visible measurements, and may contain
systematic errors as well.  This could be potentially problematic as
it would be difficult to distinguish instrumental depolarization from
an incorrect scaling assumption.

The angle of polarization would not be expected to be changed by
depolarization due to diminishing dust absorption. However, if the
intervening dust was causing a rotation in the polarization angle,
this effect would likely show a wavelength dependence. A potential
example of this is the standard HD17747, which has a literature angle
of polarization at $V$ band of $139.7^{\circ}$ but measured angles of
polarization at $J$ and $H$ bands of $125^{\circ}$ and $113^{\circ}$
respectively.

Due to these effects, visible-light polarimetric standards are
imperfect for the analysis we hope to perform, but do provide some
level of validation of the system and instrument. From our
investigation of polarized standard stars, and with the caveats above,
we can see that WIRC+Pol reproduces predicted degrees of polarization
to within $0.1\%$ as expected, though the angle of polarization
appears to show an offset of $\Delta\theta\sim-10^\circ$, with a
scatter larger than the statistical uncertainties.  These results are
consistent with the polarization accuracy and $\Delta\theta=-15^\circ$
offset found in \citep{millarblanchaer21}.

\section{Results}

We present in Table~\ref{tab.astdata} our measurements of asteroids in
the S taxonomic complex \citep[S, S/K, and Sk/Sq taxonomic types from
][]{neesePDS} and C taxonomic complex \citep[C, B, and Ch taxonomic
  types from ][]{neesePDS}.  Here we present the degree $P_r$ and
angle $\theta_r$ of the measure polarization after rotation into the
scattering plane. $P_r$ is listed as a positive value if it is
perpendicular to the scattering plane at the time of observation and
negative if it is parallel to the plane, in the typical convention for
asteroid polarimetry. The $\theta_r$ angles have not been corrected
for the offset discussed above, and after such correction are
consistent with purely parallel or perpendicular scattering for larger
polarization values.

Uncertainties on the broadband measurements include only the
statistical uncertainties of the measurements; systematic and
instrumental polarization uncertainties are $\sim0.1\%$ as described
above.  We expect that there are correlated errors between wavelength
channels, and as such the broadband uncertainties presented here may
underestimate the true uncertainty somewhat.  As our survey
progresses, the scatter on repeated measurements of the same objects
will allow us to set empirical constraints on the errors.

\begin{center}
\scriptsize
  \noindent
\begin{longtable}{cllcclc}
\caption{WIRC+Pol Asteroid Results}\label{tab.astdata}\\
\hline\hline

Asteroid & Date & Midpoint UT & Phase Angle (deg) & Filter & Polarization$~P_r^\dagger$ & Polarization Angle$~\theta_r^{\dagger\dagger}$ (deg)  \\ 

\hline\hline
\endfirsthead
\caption[]{(continued)}\\
\hline\hline
Asteroid & Date & Midpoint UT & Phase Angle (deg) & Filter & Polarization$~P_r^\dagger$ & Polarization Angle$~\theta_r^{\dagger\dagger}$ (deg)\\ 

\hline\hline
\endhead
\hline
\endfoot

\hline
\textbf{S-Complex}&&&&&\\
\hline
3 Juno & 2021-05-30 & 07:53 &  6.0 & J & $-0.74 \pm 0.02$ & $80.5 \pm 0.5$ \\ 
3 Juno & 2021-05-30 & 07:53 &  6.0 & H & $-0.88 \pm 0.01$ & $82.0 \pm 0.5$ \\ 
3 Juno & 2021-06-26 & 07:18 &  8.6 & J & $-0.89 \pm 0.01$ & $ 85.2 \pm 0.9$ \\ 
3 Juno & 2021-06-26 & 07:18 &  8.6 & H & $-0.85 \pm 0.01$ & $ 62.1 \pm 0.6$ \\ 
3 Juno & 2021-09-04 & 03:56 & 18.4 & J & $-0.46 \pm 0.04$ & $ 77.8 \pm 0.9$ \\ 
3 Juno & 2021-09-04 & 03:56 & 18.4 & H & $-0.43 \pm 0.04$ & $ 85.0 \pm 0.5$ \\ 

\hline
7 Iris & 2021-09-04 & 11:45 & 30.0 & J &  $0.81 \pm 0.03$ & $ 172.7 \pm 0.4$ \\ 
7 Iris & 2021-09-04 & 11:45 & 30.0 & H &  $0.67 \pm 0.02$ & $ 173.6 \pm 0.2$ \\ 
7 Iris & 2021-11-08 & 12:32 & 29.0 & J &  $0.74 \pm 0.04$ & $ 6 \pm 1$ \\ 
7 Iris & 2021-11-08 & 12:32 & 29.0 & H &  $0.61 \pm 0.02$ & $ 16.9 \pm 0.1$ \\ 
\hline
15 Eunomia & 2019-08-30 & 07:50 & 9.5 & J & $-0.99 \pm 0.02$ & $  84.2\pm0.2 $ \\ 
15 Eunomia & 2021-02-03 & 05:47 & 5.6 & J & $-0.89 \pm 0.01$ & $  84.2\pm0.1 $ \\ 
15 Eunomia & 2021-02-03 & 05:47 & 5.6 & H & $-0.82 \pm 0.01$ & $  89.5\pm0.1 $ \\ 
\hline
20 Massalia & 2021-11-08 & 12:01 & 28.7 & J & $0.74 \pm 0.04$ & $  169.0\pm0.5 $ \\ 
20 Massalia & 2021-11-08 & 12:01 & 28.7 & H & $0.81 \pm 0.02$ & $  170.2\pm0.2 $ \\ 
\hline
\textbf{C-Complex}&&&&&\\
\hline
1 Ceres & 2021-09-04 & 10:57 & 21.0 & J & $ 0.84 \pm 0.02$ & $157.7 \pm 0.5$ \\ 
1 Ceres & 2021-09-04 & 10:57 & 21.0 & H & $ 0.88 \pm 0.02$ & $163.8 \pm 0.2$ \\ 
1 Ceres & 2021-11-08 & 10:39 &  8.3 & J & $-1.74 \pm 0.02$ & $83.4 \pm 0.1$ \\ 
1 Ceres & 2021-11-08 & 10:39 &  8.3 & H & $-1.61 \pm 0.03$ & $86.6 \pm 0.1$ \\ 
\hline
2 Pallas & 2019-03-17 & 09:35 & 13.9 & J & $-0.97 \pm 0.01$ & $87.4 \pm 0.2$ \\ 
2 Pallas & 2019-03-17 & 10:10 & 13.9 & H & $-0.83 \pm 0.01$ & $90.3 \pm 0.2$ \\ 
2 Pallas & 2021-05-30 & 11:19 & 17.5 & J & $-0.29 \pm 0.01$ & $79 \pm 2$ \\ 
2 Pallas & 2021-05-30 & 11:19 & 17.5 & H & $-0.24 \pm 0.02$ & $109.1 \pm 0.5$ \\ 
2 Pallas & 2021-06-26 & 10:44 & 18.0 & J & $-0.15 \pm 0.03$ & $100.4 \pm 6$ \\ 
2 Pallas & 2021-06-26 & 10:44 & 18.0 & H & $-0.34 \pm 0.01$ & $115.3 \pm 0.1$ \\  
2 Pallas & 2021-09-04 & 07:32 &  3.2 & J & $-1.29 \pm 0.02$ & $83.0 \pm 0.6$ \\ 
2 Pallas & 2021-09-04 & 07:32 &  3.2 & H & $-1.22 \pm 0.01$ & $83.2 \pm 0.4$ \\ 
2 Pallas & 2021-11-08 & 04:21 & 17.4 & J & $-0.27 \pm 0.01$ & $92 \pm 6$ \\ 
2 Pallas & 2021-11-08 & 04:21 & 17.4 & H & $-0.10 \pm 0.02$ & $78 \pm 3$ \\ 
\hline
145 Adeona & 2021-11-08 & 07:02 & 6.7 & J & $-1.63 \pm 0.03$ & $84.2 \pm 0.2$  \\  
145 Adeona & 2021-11-08 & 07:02 & 6.7 & H & $-1.69 \pm 0.03$ & $85.3 \pm 0.2$ \\ 

\hline
\end{longtable}
$^\dagger$Polarization measurement $P_r$ has been rotated such that
positive values represent polarization perpendicular to the
Sun-Asteroid-Telescope scattering plane and negative values represent
polarization in the scattering plane.
\\$^{\dagger\dagger}$The angle of the polarization vector,
  rotated such that $0^\circ$ is aligned perpendicular to the
  scattering plane. 
\end{center}

In the course of our asteroid investigations it became clear that the
C-complex asteroids observed do not show any wavelength dependence on
their expected polarization.  Further, the angle of polarization on
the sky is fixed by the observing geometry and also will not have a
wavelength dependence.  As such, we can use bright C-complex asteroids
as standards to validate the measured polarization degree and angle of
WIRC+Pol.  Using asteroids as a standard has an added layer of
complexity as the expected polarization degree and on-sky angle for
these objects will be time-dependent as the viewing geometry changes.
They also cannot always be used as polarized standards, as near the
inversion angle they will be expected to be unpolarized.  These
caveats aside, the lack of wavelength dependence to both polarization
degree and angle outweighs the added planning difficulty.

\begin{figure}[ht]
\begin{center}
\includegraphics[scale=0.5]{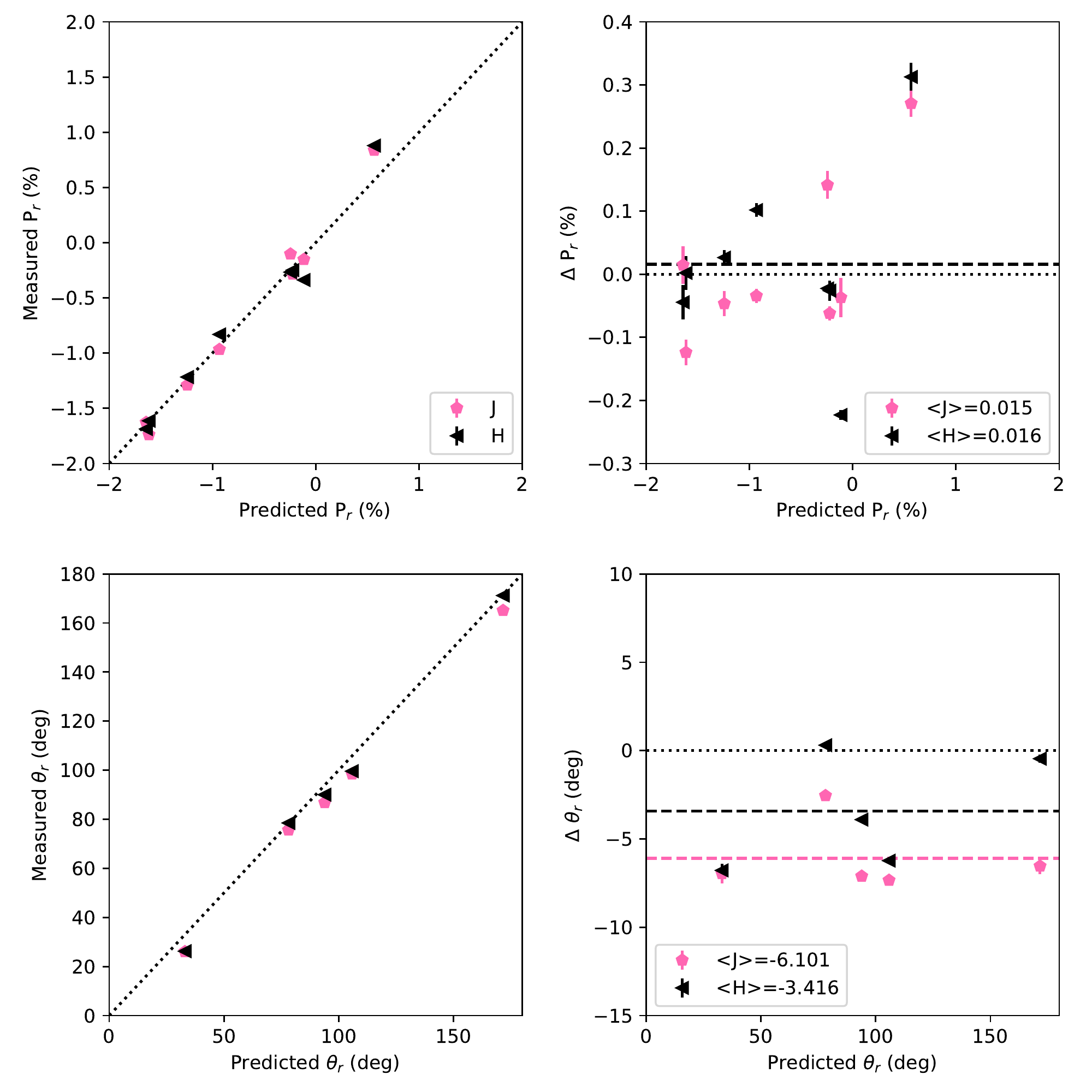}
\protect\caption{Comparison of predicted and measured polarization
  degrees and angles for C-complex asteroids, which are not expected
  to show any chromatic variations.  Plot layout is the same as shown
  in Figure~\ref{fig.pol}. Polarization degree $P$ here is given in
  the typical fashion for asteroid measurements where negative values
  indicate polarization in the scattering plane and positive values
  perpendicular to it, and no wavelength-dependent scaling has been
  applied. Polarization angles were only compared for cases where the
  predicted polarization was $|P|>0.25\%$.}
\label{fig.Ccomp}
\end{center}
\end{figure}

We show in Figure~\ref{fig.Ccomp} a comparison of our measured $J$ and
$H$ polarization degrees and angles for C-type asteroids compared to
the values predicted based on visible light polarimetric behavior
(with no chromatic correction) and the scattering geometry at the time
of observation.  Angle of polarization $\theta_r$ was only compared for
observing dates where the predicted polarization degree was expected
to be greater than $|P_r|>0.25$ to ensure that the angle measurements
were significant.  While the number of observations are currently
limited, our initial results indicate that any systematic instrumental
polarization is well below the $0.1\%$ level, that the measurement
scatter is consistent with the $0.1\%$ expected value, and that there
is an offset of $\Delta\theta\sim-5^\circ$.  This offset is somewhat
lower than the $\theta$ offset seen for the polarized stars in this
work or the offset calculated by \citet{millarblanchaer21}.  Future
observations of C-type asteroids will refine this measurement.

\clearpage

\section{Polarimetric phase curves}

To search for changes in polarization behavior with wavelength, we
combined our measurements obtained in the near-infrared $J$ and $H$
bands with literature measurements of shorter wavelength polarization
data that are found in the Asteroid Polarimetric Database \citep{APD}.
For this initial investigation we combined data from multiple
different asteroids of the same taxonomic complex to allow us to
search for broader trends; future observations will seek to obtain
full phase curves for individual objects, allowing for
object-by-object comparisons as well.

\begin{figure}[ht]
\begin{center}
\includegraphics[scale=0.7]{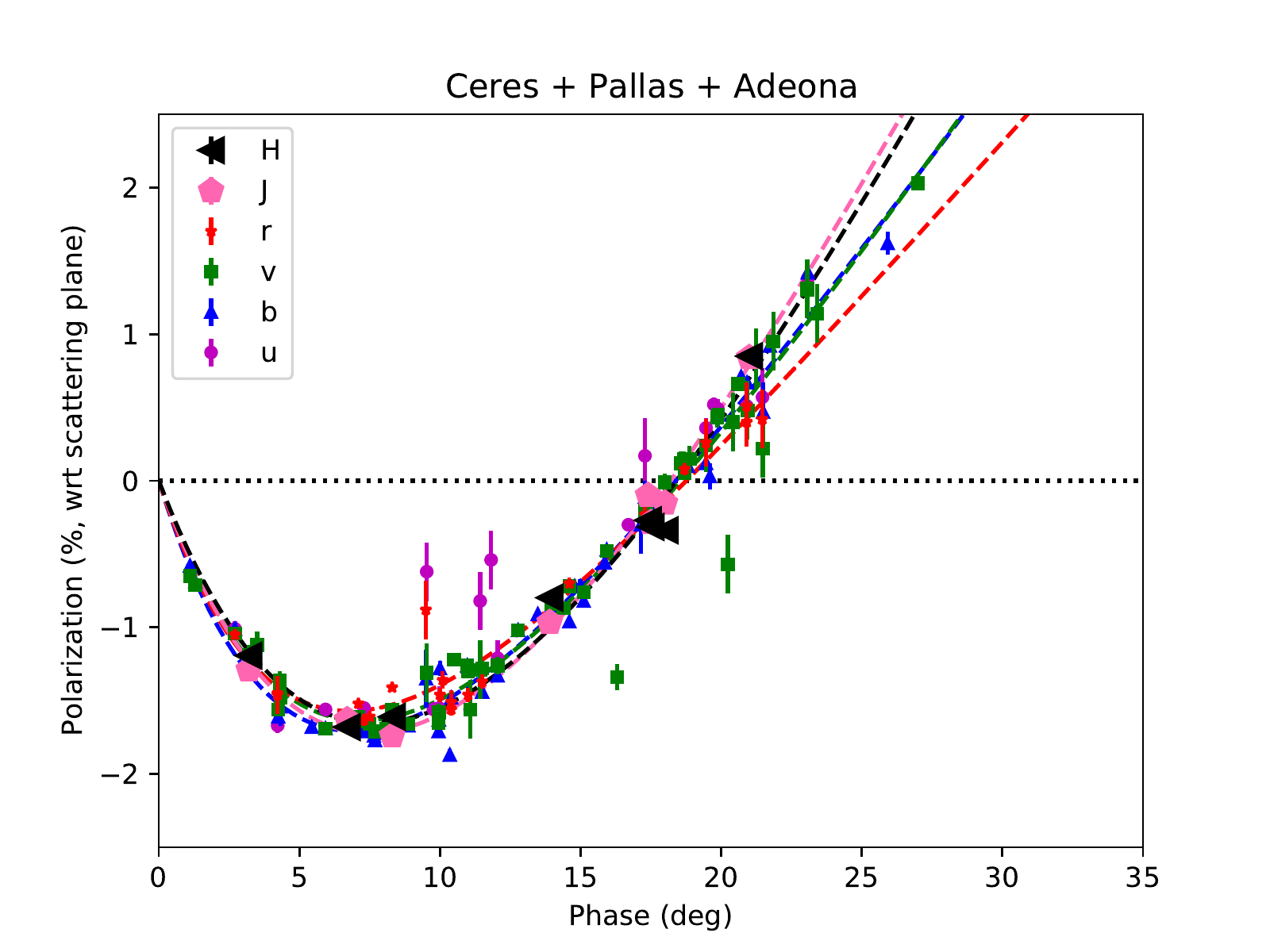}
\protect\caption{Polarimetric-phase curve for three C-complex
  asteroids: Ceres, Pallas, and Adeona. $J$ and $H$ band data are
  presented in this manuscript, while $u$, $b$, $v$, and $r$ band data
  for these three objects are from the Asteroid Polarimetric Database
  \citep{APD}. Functional form fits to the data using
  Eq~\ref{eq.curve} are shown in dashed lines, with the color of the
  line corresponding to the bandpass being fit.}
\label{fig.Cpol}
\end{center}
\end{figure}

\begin{figure}[ht]
\begin{center}
\includegraphics[scale=0.7]{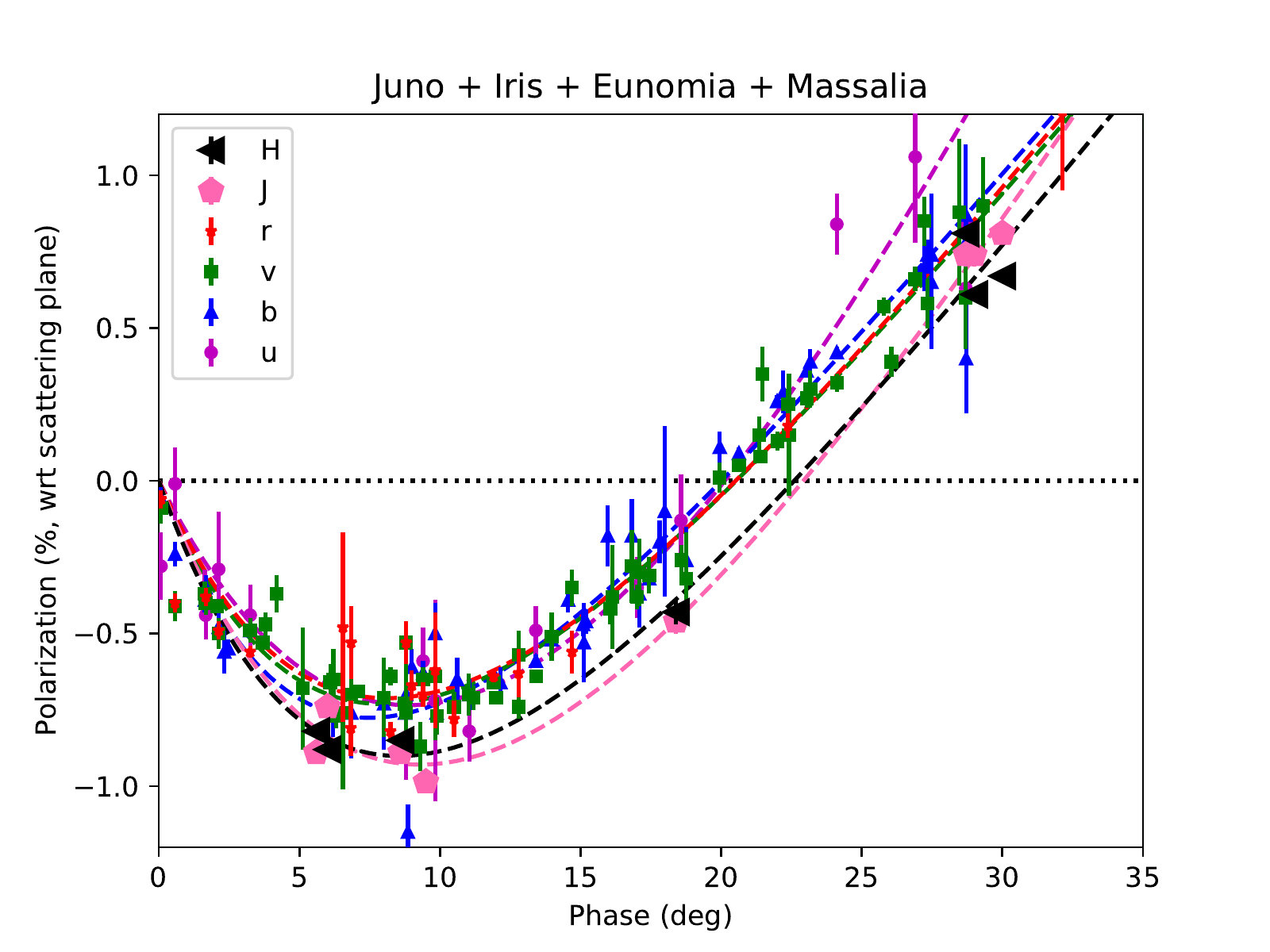}
\protect\caption{Polarimetric-phase curve for four S-complex
  asteroids: Juno, Iris, Eunomia, and Massalia. $J$ and $H$ band data are
  presented in this manuscript, while $u$, $b$, $v$, and $r$ band data
  for these three objects are from the Asteroid Polarimetric Database
  \citep{APD}. Functional form fits to the data using
  Eq~\ref{eq.curve} are shown in dashed lines, with the color of the
  line corresponding to the bandpass being fit.}
\label{fig.Spol}
\end{center}
\end{figure}

We show our results for the C-complex asteroids in
Figure~\ref{fig.Cpol}, and for the S-complex asteroids in
Figure~\ref{fig.Spol}.  The change in polarization as a function of
asteroid phase can be described by the equation from
\citet{muinonen09}:

\begin{equation}
P=A~\left(e^{-\alpha/d} - 1\right) + k\alpha
  \label{eq.curve}
  \end{equation}

\noindent where $A$ is an amplitude parameter, $k$ is a slope
parameter, and $d$ is the width of the negative branch.  By fitting
this function to the polarimetric measurements we can derive the phase
curve parameters $P_{min}$, $\alpha_{min}$, $h$, and $\alpha_0$ that
are related to asteroid physical properties.  The fitted parameters
and derived phase curve parameters for the S- and C-complex objects
are given in Table~\ref{tab.fits}, along with the 16$^{th}$ and
84$^{th}$ percentile uncertainty ranges.

\begin{center}
\scriptsize
  \noindent
\begin{longtable}{lccccccc}
\caption{Asteroid Polarization Phase Curve Fits}\label{tab.fits}\\
\hline\hline

Band & $A$ & $k$ & $d$ & $P_{min}$ ($\%$) & $\alpha_{min}$ (deg) & $h$ & $\alpha_0$ (deg) \\

\hline\hline
\endfirsthead
\caption[]{(continued)}\\
\hline\hline
Band & $A$ & $k$ & $d$ & $P_{min}$ ($\%$) & $\alpha_{min}$ (deg) & $h$ & $\alpha_0$ (deg) \\

\hline\hline
\endhead
\hline
\endfoot

\hline
\textbf{S-Complex}&&&&&\\
\hline
u & 6$^{+5}_{-2}$ & 0.21$^{+0.1}_{-0.05}$ & 14$^{+9}_{-4}$ & -0.75$^{+0.08}_{-0.06}$ & 8.8$^{+0.7}_{-0.7}$ & 0.116$^{+0.007}_{-0.007}$ & 20.1$^{+0.7}_{-0.5}$ \\
b & 2.2$^{+0.2}_{-0.2}$ & 0.108$^{+0.007}_{-0.007}$ & 6.2 $^{+0.7}_{-0.6}$ & -0.78$^{+0.02}_{-0.02}$ & 7.5$^{+0.3}_{-0.3}$ & 0.094$^{+0.003}_{-0.004}$ & 20.0$^{+0.2}_{-0.2}$ \\
v & 2.4$^{+0.2}_{-0.2}$ & 0.109$^{+0.009}_{-0.006}$ & 7.1 $^{+0.7}_{-0.6}$ & -0.73$^{+0.02}_{-0.01}$ & 7.9$^{+0.2}_{-0.2}$ & 0.091$^{+0.02}_{-0.03}$ & 20.6$^{+0.1}_{-0.2}$ \\
r & 2.5$^{+1.5}_{-0.7}$ & 0.12$^{+0.05}_{-0.03}$ & 8 $^{+4}_{-2}$ & -0.71$^{+0.03}_{-0.04}$ & 8.1$^{+0.9}_{-0.7}$ & 0.09$^{+0.01}_{-0.01}$ & 20.8$^{+0.8}_{-0.7}$ \\
J & 3.8$^{+0.4}_{-0.5}$ & 0.15$^{+0.01}_{-0.01}$ & 10 $^{+1}_{-1}$ & -0.93$^{+0.01}_{-0.01}$ & 9.4$^{+0.2}_{-0.3}$ & 0.112$^{+0.003}_{-0.004}$ & 23.0$^{+0.2}_{-0.3}$ \\
H & 2.7$^{+0.3}_{-0.2}$ & 0.113$^{+0.009}_{-0.006}$ & 7.1 $^{+0.9}_{-0.5}$ & -0.90$^{+0.1}_{-0.1}$ & 8.5$^{+0.3}_{-0.3}$ & 0.098$^{+0.004}_{-0.003}$ & 22.6$^{+0.3}_{-0.3}$ \\

\hline
\textbf{C-Complex}&&&&&\\
\hline

b & 4.9$^{+0.2}_{-0.1}$ & 0.26$^{+0.01}_{-0.01}$ & 5.6$^{+0.3}_{-0.2}$ & -1.71$^{+0.01}_{-0.01}$ & 6.88$^{+0.1}_{-0.07}$ & 0.225$^{+0.004}_{-0.003}$ & 18.4$^{+0.1}_{-0.1}$ \\
v & 5.4$^{+0.3}_{-0.2}$ & 0.28$^{+0.01}_{-0.01}$ & 6.6$^{+0.3}_{-0.3}$ & -1.63$^{+0.01}_{-0.02}$ & 7.22$^{+0.09}_{-0.09}$ & 0.226$^{+0.005}_{-0.005}$ & 18.6$^{+0.1}_{-0.1}$ \\
r & 4.1$^{+0.4}_{-0.4}$ & 0.22$^{+0.02}_{-0.02}$ & 5.1$^{+0.5}_{-0.5}$ & -1.58$^{+0.03}_{-0.02}$ & 6.7$^{+0.2}_{-0.2}$ & 0.19$^{+0.01}_{-0.01}$ & 18.7$^{+0.3}_{-0.2}$ \\
J & 7.5$^{+0.4}_{-0.4}$ & 0.37$^{+0.02}_{-0.01}$ & 8.3$^{+0.5}_{-0.4}$ & -1.72$^{+0.01}_{-0.02}$ & 7.50$^{+0.07}_{-0.08}$ & 0.265$^{+0.005}_{-0.003}$ & 18.2$^{+0.04}_{-0.04}$ \\
H & 7.8$^{+0.5}_{-0.5}$ & 0.37$^{+0.02}_{-0.02}$ & 9.0$^{+0.5}_{-0.5}$ & -1.65$^{+0.02}_{-0.02}$ & 7.67$^{+0.07}_{-0.06}$ & 0.257$^{+0.004}_{-0.005}$ & 18.4$^{+0.02}_{-0.02}$ \\

\hline
\end{longtable}
\end{center}

We find that the C-complex objects do not show any significant change
in their best-fit polarimetric phase curve at the phase angles
investigated here from $B$ band to $H$ band, a wavelength span of over
$1~\mu m$.  This is somewhat different than what was found by
\citet{belskaya09}, who find an expected change of $+0.5\%$ at
$\alpha=20^\circ$, though it is consistent with their results for
Ceres alone, which showed nearly no change over their bandpasses.
This may indicate that object size is important for dictating
wavelength dependence.  This will be investigated further as we
continue our survey of asteroid NIR polarimetric properties.

Conversely to what we see for the C-complex, for the S-complex objects
we find that the phase curve appears to be offset or stretched to
higher phase angles at longer wavelengths.  This is comparable to what
was observed by \citet{belskaya09} in direction and magnitude, though
we find a change between $r$ and $J$ bands while they found a change
from $U$ to $i$.  From the results shown in \citet{masiero09}, this
behavior is what would be expected for a change in the index of
refraction $n$ of the surface material. The interplay between
  the changing wavelength of light and the size distribution of
  scattering elements could also be a contributing factor in the
  observed differences.  Using the relationship described by
\citet{gilhutton17}:
\[n = (0.0403 \pm 0.0082) \alpha_0 + (0.9438 \pm 0.1650)\]
we can infer that the magnitude of the expected index of refraction
change would be $\Delta n\sim0.09$ from R-band to J-band if the
observed shift was due entirely to changing $n$.

Interestingly, the $h$ slope values and $P_{min}$ are comparable
across all bands.  If the fitted constants in the slope-albedo
relationship \citep[e.g.][]{cellino15} are not wavelength-dependent
then this would imply the geometric albedos are the same across
wavelengths.  These findings are preliminary though, and full phase
curve coverage for individual objects will help untangle any effects
due to differences in surface composition between objects within the
S-complex.

\section{Conclusions}

We have presented initial results of our survey of asteroid
polarimetric properties in the near-infrared using the WIRC+Pol
instrument on the Palomar 200-inch telescope.  Our measurements of
standard stars confirm that the instrument is capable of providing
absolute polarimetric sensitivity at the $0.1\%$ level in both $J$ and
$H$ bands.  Polarization of starlight by intervening dust is strongly
wavelength dependent, complicating this verification, but our
observations of the polarization of C-complex asteroids indicates that
these objects have reliable polarization degrees and angles from
visible to near-infrared wavelengths, and could be used for
verification of the calibration of near-infrared polarimeters. Further
polarimetric observations of C-complex asteroids in the near-infrared
are encouraged to ensure that there are no systematic issues arising
from use of these objects as calibration targets.

We also presented phase curves for seven S-complex and C-complex
asteroids.  C-complex asteroids show no significant changes in their
polarization-phase curves as a function of wavelength. In contrast,
S-complex objects do show a significant change, most notable as an
increase in polarimetric inversion angle.  This may be linked to a
change in the index of refraction of the surface material as a
function of wavelength.

Our initial findings demonstrate the need for a more complete survey
of the near-infrared polarimetric properties of different asteroid
taxonomies.  A specific focus on obtaining full phase curve coverage
for individual objects will help disentangle changes among objects
within a spectral class from differences between classes.  Future
WIRC+Pol observations will allow us to build a more complete picture
of the physical properties and mineral makeup of the small bodies of
our Solar system.

\section*{Acknowledgments}

The authors would like to thank the Palomar staff for their dedicated
support and assistance that enabled our WIRC+Pol observing runs,
especially as the COVID-19 pandemic up-ended regular operating
procedures.  We also thank the referees for their comments and
suggestions that improved this paper.  Based on observations obtained
at the Hale Telescope, Palomar Observatory as part of a continuing
collaboration between the California Institute of Technology,
NASA/JPL, Yale University, and the National Astronomical Observatories
of China.  This research made use of Photutils, an Astropy package for
detection and photometry of astronomical sources (Bradley et al.
2019).


\begin{thebibliography}{XXX}

\bibitem[Belskaya \etal(2009)]{belskaya09}
  Belskaya, I.N., Levasseur-Regourd, A.-C., Cellino, A., \etal, 2009, Icarus, 199, 97.


\bibitem[Bradley \etal(2019)]{bradley19}
  Bradley, L. \etal, 2019, astropy/photutils 0.7.2, Zenodo, doi:10.5281/zenodo.3568287

\bibitem[Ca\~{n}ada-Assandri \etal(2012)]{canada12}
Ca\~{n}ada-Assandri, M., Gil-Hutton, R., \& Benavidez, 2012, A\&A, 542, 11.
  
\bibitem[Cellino \etal(2005a)]{cellino05}
Cellino, A., Gil Hutton, R., di Martino, M., Bendjoya, Ph., Belskaya, I.N. \& Tedesco, E.F., 2005, Icarus, 179, 304.

\bibitem[Cellino \etal(2015)]{cellino15}
Cellino, A., Bagnulo, S., Gil-Hutton, R. \etal, 2015, MNRAS, 451, 3473.

\bibitem[Gil-Hutton \& Ca\~{n}ada-Assandri(2011)]{gilhutton11}
Gil-Hutton, R. \& Ca\~{n}ada-Assandri, M., 2011, A\&A, 529, 86.

\bibitem[Gil-Hutton \& Ca\~{n}ada-Assandri(2012)]{gilhutton12}
Gil-Hutton, R. \& Ca\~{n}ada-Assandri, M., 2012, A\&A, 539, 115.


\bibitem[Gil-Hutton \& Garc\'{i}a-Migani(2017)]{gilhutton17}
Gil-Hutton, R. \& Garc\'{i}a-Migani, E., 2017, A\&A, 607, 103.

\bibitem[Heiles(2000)]{heiles00}
  Heiles, C., 2000, AJ, 119, 923.

\bibitem[Hopfield(1966)]{hopfield66}
  Hopfield, J.J., 1966, Science, 151, 1380.


\bibitem[Jones \etal(2008)]{jones08}
  Jones, T.,J., Stark, D., Woodward, C., \etal, 2008, AJ, 135, 1318. 
  
\bibitem[Kwon \etal(2019)]{kwon19}
  Kwon, Y., Ishiguro, M., Kwon, J., \etal, 2019, A\&A, 629, 121.
  
\bibitem[Lupishko(2022)]{APD}
  Lupishko, D., Ed. (2022). Asteroid Polarimetric Database V2.0. urn:nasa:pds:asteroid\_polarimetric\_database::2.0. NASA Planetary Data System

\bibitem[Lyot(1929)]{lyot29}
  Lyot, B., 1929, Ann. Obs. Meudon, 8, 1.

\bibitem[Masiero \etal(2009)]{masiero09}
Masiero, J., Hartzell, C., Scheeres, D., 2009, AJ, 138, 1557.

\bibitem[Millar-Blanchaer \etal(2021)]{millarblanchaer21}
  Millar-Blanchaer, M., Tinyanont, S., Jovanovic, N., \etal, 2021, SPIE, 11447, 114475Y.

\bibitem[Muinonen \etal(2002)]{muinonen02}
Muinonen, K., Piironen, J., Shkuratov, Y., Ovcharenko, A. \& Clark, B., 2002, Asteroids III, ed. Bottke, Cellino, Paolicchi \& Binzel (Univ of Arizona Press), 123.

\bibitem[Muinonen \etal(2009)]{muinonen09}
Muinonen, K., Penttil\"{a}, A., Cellino, A., \etal, 2009, M\&PS, 44, 1937.

\bibitem[Neese(2017)]{neesePDS}
Neese, C., Ed., 2017, Asteroid Taxonomy V1.0. urn:nasa:pds:ast\_taxonomy::1.0. NASA Planetary Data System; https://doi.org/10.26033/e1p3-xm59.

\bibitem[Oishi \etal(1978)]{oishi78}
  Oishi, M., Kawara, K., Kobayashi, Y., \etal, 1978, PASJ, 30, 149.

\bibitem[Pan \& Ip(2022)]{pan22}
  Pan, K.-S. \& Ip, W.-H., 2022, PSS, 212, 105412.

\bibitem[Serkowski \etal(1975)]{serkowski75}
  Serkowski, K., Mathewson, D.S., \& Ford, V.L., 1975, ApJ, 196, 261.

\bibitem[Shkuratov \etal(1994)]{shkuratov94}
  Shkuratov, Yu., Muinonen, K., Bowell, E., \etal, 1994, EM\&P, 65, 201. 
  
\bibitem[Tinyanont \etal(2019a)]{tinyanont19a}
  Tinyanont, S., Millar-Blanchaer, M.A., Nilsson, R., \etal 2019a, PASP, 131, 25001.

\bibitem[Tinyanont \etal(2019b)]{tinyanont19b}
  Tinyanont, S., Millar-Blanchaer, M.A., Jovanovic, N. \etal 2019b, Proc. SPIE, 11132, 1113209.


\bibitem[Wilson \etal(2003)]{wilson03}
  Wilson, J.~C., Eikenberry, S.~S., Henderson, C.~P., \etal 2003, Proc. SPIE, 4841, 451.

\bibitem[Zellner \etal(1974)]{zellner74}
Zellner, B., Gehrels, T. \& Gradie, J., 1974, AJ, 79, 1100.

\bibitem[Zellner \& Gradie(1976)]{zellner76}
Zellner, B. \& Gradie, J., 1976, AJ, 81, 262.

\end{thebibliography}
\end{document}